# Metallic nature and site-selective magnetic collapse in iron oxide $Fe_4O_5$ at the extreme conditions of Earth's deep interior

Aiqin Yang[1][#], Qiaoying Qin[1][#], Xiangru Tao[1], Shengli Zhang[1], Yongtao Zhao[1][*], Peng Zhang[1][*]

[1]MOE Key Laboratory for Nonequilibrium Synthesis and Modulation of Condensed Matter, School of Physics, Xi'an Jiaotong University, Xi'an, 710049, China
#The two authors contribute equally to this work.
*Corresponding authors: zhaoyongtao@xjtu.edu.cn (Y. T. Zhao)
zpantz@xjtu.edu.cn (P. Zhang).

Properties of iron oxides at the extreme conditions are of essential importance in condensed matter physics and Geophysics. The recent discovery of a new type of iron oxide, $Fe_4O_5$, at high pressure and high temperature of Earth's deep interior attracts great interests. In this paper, we report the electronic structure and the magnetic properties of $Fe_4O_5$ predicted by the density functional theory plus dynamic mean field theory (DFT+DMFT) approach. We find that $Fe_4O_5$ stays metallic from ambient pressure to high pressure. The magnetic moments of iron atoms at the three different crystallographic positions of $Fe_4O_5$ undergo position-dependent collapse as being compressed. Such site-selective magnetic moment collapse originates from the shift of energy levels and the consequent charge transfer among the Fe-3$d$ orbits under compression.

**PACS Codes:** 91.60.Gf; 91.60.Pn

## 1 INTRODUCTION

Iron oxides that consist of the two most abundant elements on Earth are important components of Earth's mantle [1-8]. They account for about 7.5% of Earth's mass, only next to silicon and magnesium oxides. The electronic and the thermal conductivity of iron oxides are crucial for modeling the mantle of Earth and the interiors of other planets with compositions of iron and oxygen [9-11]. Iron oxides have also been widely used in industry [12,13] and biomedicine [14]. Therefore their properties have long attracted extensive interests of scientific communities.

Earth's mantle is in a high temperature, high pressure state. Typically, the temperature ranges from 500 K to 3000 K and the pressure ranges from 20 GPa to 140 GPa. At such extreme conditions, the oxidation states, the spin states and the phase stabilities of iron oxides may alter according to the environment [15,16]. The exploration of their properties at Earth's deep interior is imperative to understand the seismic observation and dynamic processes of Earth's mantle and outer core [3,7,8,17-19]. At the ambient conditions, there are three known forms of iron oxides including wüstite (FeO), magnetite ($Fe_3O_4$), hematite ($Fe_2O_3$) [2,7,12,13,15,19,20]. Recently, a series of new iron oxides were identified at high pressure and high temperature including $FeO_2$ [5,8,19,21], $Fe_4O_5$ [9,15], $Fe_5O_6$ [4,22], $Fe_5O_7$ [21], $Fe_7O_9$ [23] and $Fe_{13}O_{19}$ [24]. In 2011, Lavina *et al.* firstly reported the synthesis of a stoichiometric iron oxide $Fe_4O_5$ at 10 GPa and 1800K [15]. $Fe_4O_5$ possesses a $CaFe_3O_5$-type crystal structure with space group *Cmcm* [3,7,15,25]. Its atomic arrangement can be viewed as a stacking along the *c*-axis of layers of edge-sharing $FeO_6$ octahedra with two slightly different iron positions Fe1 (4*a* site) and Fe2 (8*f* site), and layers of face-sharing trigonal prisms with the third iron position Fe3 (4*c* site) (see Fig.1).

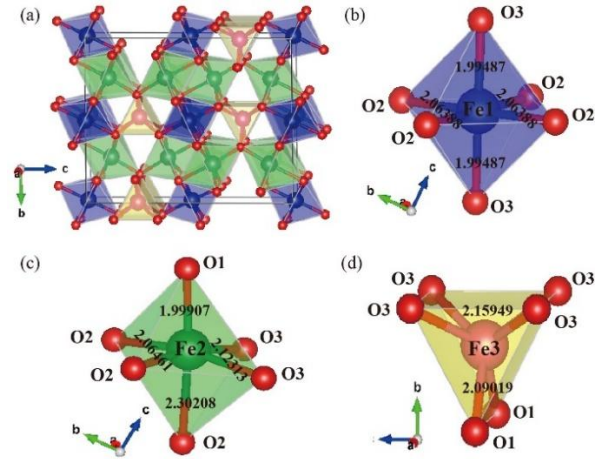

Fig.1 (Color online.) Crystal structure of $Fe_4O_5$ (a), $FeO_6$ octahedra (b-c) and $FeO_6$ trigonal prisms (d). The blue and green octahedra show the local coordination of sites Fe1 (4*a*) and Fe2 (8*f*) and the yellow trigonal prism shows the local coordination of site Fe3 (4*c*). The bond lengths at the ambient conditions are marked in the subfigures.

The discovery of $Fe_4O_5$ immediately stimulates great interests in exploring the properties of this new iron oxide since it is energetically more favorable with increased pressures than stoichiometric $Fe_3O_4$ and FeO [2]. This opens up the possibility that $Fe_4O_5$ could be stable in Earth's mantle. Density functional theory calculations suggest an antiferromagnetic ground state for $Fe_4O_5$ [3]. Experiments by Ovsyannikov *et al.* indicate an unusual charge-ordering transition in $Fe_4O_5$ at about 150K, resulting from competing formation of Fe dimer and trimer [25]. Later they further determined the phase diagram of $Fe_4O_5$ at high pressures and low temperatures, pointing out that the charge ordering can be tuned by pressure [7]. However, the investigation of $Fe_4O_5$ at high pressures and high temperatures, which is of critical importance for the modeling of Earth's mantle, is totally absent. In this letter we will present a DFT+DMFT [26] calculation of the electronic and the magnetic properties of the strongly correlated material $Fe_4O_5$ at the high temperature and the high pressure conditions of Earth's deep interior.

## 2 COMPUTATIONAL METHODS

*The DFT+DMFT method.* In DFT+DMFT calculation, first we adopt the all electron LAPW Wien2k package [27] in the DFT part to get converged charge density, eigenvectors and eigenenergies of the Kohn-Sham equations. We choose the Wu-Cohen [28] exchange correlation function and a 20×20×4 k-point mesh. The cutoff energy is -6.0 Ry and $R_{mt} \cdot K_{max} = 7$, where $K_{max}$ is the magnitude of the largest $K$ vector and $R_{mt}$ is the smallest atomic sphere radius in the unit cell. The smallest atomic radius of Fe is 1.65 bohr and that of O is 1.42 bohr at the 45% volume compression. The DFT convergence criteria of charge and energy are set as $1×10^{-5}$ e and $1×10^{-5}$ Ry respectively. In the DMFT [29] part, the hybridization expansion continuous time quantum Monte Carlo (CTQMC) [30,31] is used as the impurity solver of the DMFT self-consistent equations to get the self-energy. A total number of $10^7$ Monte Carlo updates are used in the DMFT iteration. After a DMFT iterations, the charge density is updated and then being used to construct the new Kohn-Sham potential for the next DFT iterations. In the Supplementary Information Fig. S1-S2, we presented the evolution of the self-energy and the local Green's function with respect to the DMFT iterations for the five Fe-3$d$ orbits of the three iron atoms at 300 K and at the ambient pressure. After 30 DMFT iterations, the DFT+DMFT calculations stop when we have reached fully convergence of the self-energy and the Green's function. The double counting energy is calculated from the fully localized limit method [32]. The DFT+DMFT calculated self-energy on the imaginary axis is rotated to the real axis by the maximum entropy analytical continuation method [33].

*Coulomb interaction U and Hund's coupling J.* In the CTQMC part, the requested screened Coulomb interaction U and the Hund's coupling J of the Fe-3$d$ orbitals are determined by the constrained density functional theory (cDFT) method [34]. There are three non-equivalent iron atoms Fe1 (4$a$ site), Fe2 (8$f$ site) and Fe3 (4$c$ site) in $CaFe_3O_5$-type $Fe_4O_5$. We calculated the Coulomb U and the Hund's J at both $V_0 = 603.86$ bohr$^3$/$Fe_4O_5$ (volume per formula unit at the ambient pressure, a = 5.473443 bohr, b = 18.561956 bohr, c = 23.774408 bohr) and V = 572.13 bohr$^3$/$Fe_4O_5$ (in later text note as (1-V/V$_0$)% = 5.25% compression). At the ambient pressure (U, J) = (6.81 eV, 0.89 eV) at Fe1, (6.42 eV, 0.9 eV) at Fe2, and (6.91 eV, 0.94 eV) at Fe3. At 5.25% compression (U, J) = (6.53 eV, 0.88 eV) at Fe1, (6.35 eV, 0.9 eV) at Fe2, and (6.82 eV, 0.94 eV) at Fe3. Since the value of the Coulomb interaction U and the Hund's coupling J is not very sensitive to position and compression, we choose (U, J) = (6.7 eV, 0.9 eV) throughout our calculations at every volume and temperature for all Fe-3$d$ electrons. We also tested U = 5 eV and 10 eV, J = 0.7 eV and 1.12 eV at the ambient pressure. The different values of U, J only slightly change the derived charge density, density of states and the magnetic moments, thus all conclusions we presented in this letter stay valid.

*The pressure-volume relationship.* Based on our DFT+DMFT calculation of the energy-volume relationship, we derived the pressure-volume dependence of $Fe_4O_5$ at 300 K as presented in the Supplementary Information Fig. S3. Our numerical curve reasonably overlaps the powder x-ray diffraction data by Lavina *et al.* [15] below 30 GPa.

## 3 RESULTS AND DISCUSSIONS

At the ambient conditions, some iron oxides including FeO and $Fe_2O_3$ are Mott-type insulators, while magnetite $Fe_3O_4$ is metal. Our DFT+DMFT calculations indicate that the electronic conduction property of $Fe_4O_5$ is close to $Fe_3O_4$. The total and the orbital-projected density of state (DOS) of $Fe_4O_5$ are presented in Fig. 2. In Fig. 2(a) and Fig. 2(b), the peak of the DOS at the Fermi level $E_F$ clearly reveals the metallic nature of $Fe_4O_5$ from the ambient pressure to high pressures. As shown in Fig. 2(c) and Fig. 2(d), the total DOS of $Fe_4O_5$ at the Fermi level $E_F$ has dominant Fe-3$d$ character and minor O-p character that suggests the conduction in $Fe_4O_5$ is mainly due to the Fe-3$d$ electrons. In Fig. 2(c) the sharp excitation peak at about 0.2 eV above $E_F$ originates from the Kondo scattering of the itinerant O-$p$ electrons by the heavy Fe-3$d$ electrons. In the Supplementary Information Fig. S4 we present the hybridization function among the Fe-3$d$ electrons and the conduction bath at 300 K and at the ambient pressure. The hybridization peaks at ($E_F$ + 0.2) eV for each Fe-3$d$ orbits provide clear evidence of the strong hybridization among the conduction electrons and the Fe-3$d$ electrons that form the local magnetic moments. Such Kondo scattering is also observed in other transition metal oxides like FeO, MnO, CoO and NiO [35-37]. At 45% volume compression in Fig. 2(d) the peak is lower due to the band broadening.

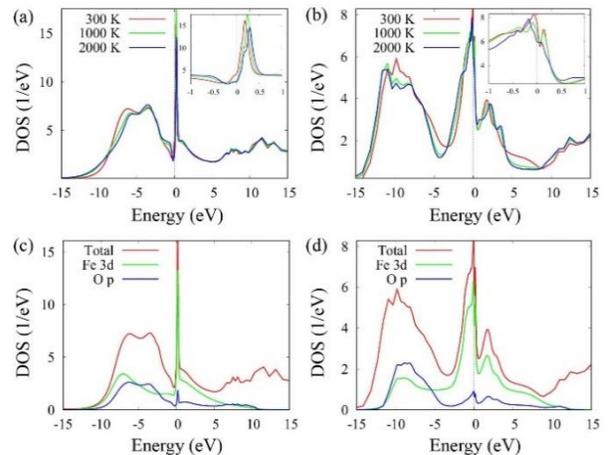

Fig. 2 (Color online.) The total DOS of $Fe_4O_5$ at the ambient pressure (a) and at 45% volume compression (b) at temperatures of 300 K, 1000 K and 2000 K. The insets show the DOS of $Fe_4O_5$ in energy window [-1 eV, 1 eV]. The orbital-projected DOS of $Fe_4O_5$ at 300 K, the ambient pressure (c) and at 300 K, 45% volume compression (d).

Previous first-principles calculations suggest magnetic moments of Fe atoms in range of ~3.6–3.8 $\mu_B$/Fe in the pressure range of 0-30 GPa [3]. However, our DFT+DMFT calculations indicate that the local magnetic moments $\sqrt{\langle m_z^2 \rangle}$ in Fe$_4$O$_5$ undergo site-selective magnetic moment collapse under compression as shown in Fig. 3. In Fig. 3, the magnetic moments of all three iron atoms are around 3.8-4.0 $\mu_B$/Fe at the ambient pressure. As Fe$_4$O$_5$ is compressed, the magnetic moment of Fe1 (4$a$) decreases firstly, then that of Fe2 (8$f$) decreases secondly, and that of Fe3 (4$c$) decreases lastly. At 45% compression endpoint, the magnetic moments of Fe1 (4$a$) and Fe2 (8$f$) saturate at about 1.0 $\mu_B$/Fe, while the magnetic moment of Fe3 (4$c$) is about 2.5 $\mu_B$/Fe. In addition, we also noticed that the magnetic moment of Fe1 (4$a$) at 300 K decreases at relative higher speed than that at 2000 K. Such temperature tuned speed of magnetic moment collapse is not that obvious at Fe2 (8$f$) and seems disappear at Fe3 (4$c$). The evolution of the average magnetic moment of iron atoms at the three sites as a function of pressure and temperature is summarized in Fig. 4. Although in Fig. 4 the temperature effect is less significant, we still find that the speed of magnetic moment collapse gradually slow down as raising temperatures, especially in the red color region. Similar site-selective magnetic moment collapse has also been observed in Fe$_5$O$_6$ [22].

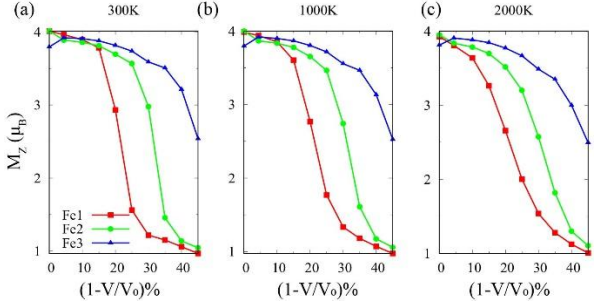

Fig. 3 (Color online.) The magnetic moments of Fe$_4$O$_5$ at 300 K (a), 1000 K (b) and 2000 K (c) as a function of compression. Red square represents the magnetic moment of Fe1 (4$a$), the green circle represents the magnetic moment of Fe2 (8$f$), and the blue triangle represents the magnetic moment of Fe3 (4$c$). (1-V/V$_0$)% is the notation of volume compression percentage, where V$_0$ = 603.86 bohr$^3$/Fe$_4$O$_5$ is the volume per formula unit at the ambient pressure.

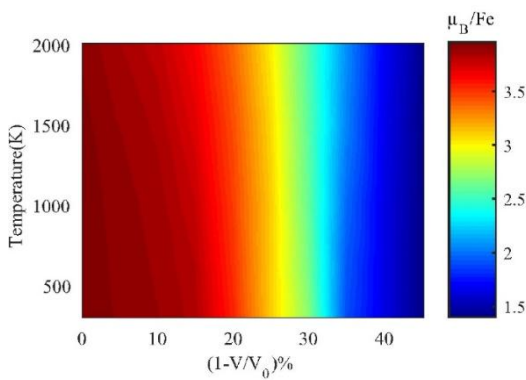

Fig. 4 (Color online.) The average magnetic moment of iron versus temperature and pressure in Fe$_4$O$_5$. Color indicates the value of magnetic moment in unit of $\mu_B$/Fe.

In transition metal oxides like FeO, Fe$_2$O$_3$ and MnO their high spin-low spin transitions coexist with the insulator-metal transitions [38-40]. Thus the insulator-metal transitions in these materials are often presumed originate from the spin fluctuations. Here in metallic Fe$_4$O$_5$ we find the site-selective spin transition could be independent of the electronic insulator-metal transition. Such behavior has also been observed in FeO$_2$, where they find a spin transition at about 14 GPa in contrast FeO$_2$ always stay metallic without pressure driven insulator-metal transition [41].

Previous works on transition metal mono-oxides prove that the collapse of magnetic moment is related to the partition of electrons at the $d$-orbit in transition metals [40,42]. Here we calculate the electron occupancies on the five Fe-3$d$ orbits at three crystallographic sites Fe1 (4$a$), Fe2 (8$f$) and Fe3 (4$c$) at 300 K. As shown in Figure 5, at Fe1 (4$a$) and Fe2 (8$f$) the electron occupancies of the $z^2$, $x^2-y^2$ and $xz$ orbits increase at compression of volume, while the electron occupancies of the $yz$ and $xy$ orbits decrease. It indicates the transfer of electrons from the $yz$ and $xy$ orbits to the $z^2$, $x^2-y^2$ and $xz$ orbits of iron at compression. However, the change of the electron occupancies at the 4$c$ position is less significant, which indicates that the electronic correlations of Fe3 atom are stronger relative to the electronic correlations of Fe1 and Fe2. The analysis of the bond lengths of the iron atoms with the surrounding oxygen atoms at three crystallographic positions confirms this view. As shown in Figure 1, the Fe-O bonds in the trigonal prism around Fe3 (4$c$) are longer than the Fe-O bonds in the octahedra around Fe1 (4$a$) and Fe2 (8$f$), indicating that the electrons in the trigonal prism are more localized.

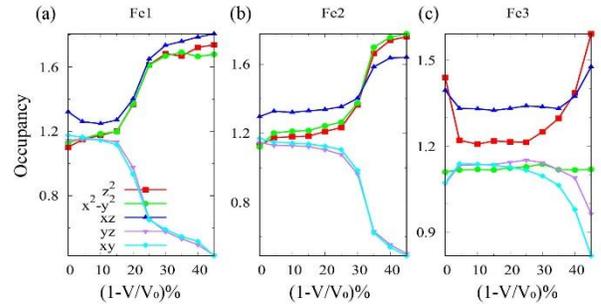

Fig. 5 (Color online.) The compression dependent electron occupancies of five Fe-3$d$ orbits at 300 K at Fe1 (4$a$), Fe2 (8$f$), and Fe3 (4$c$) in panels (a)-(c) respectively.

Such electron partitioning incurred magnetic moment collapse scenario can be understood in a simplified picture. At the ambient condition all five Fe-3$d$ orbits are almost evenly occupied and the Hund's rule maximized the magnetic moments at around 4.0 $\mu_B$/Fe. At the extreme compression condition imagining that all electrons of the $yz$ and $xy$ orbits have been transferred to the $z^2$, $x^2-y^2$ and $xz$ orbits, then the six Fe-3$d$ electrons must be pairing in singlets due to the Pauli exclusion principle and the total magnetic moment is zero.

The transfer of electrons among the five Fe-3$d$ orbits originates from the shift of energy levels of these orbits under compression. The energy difference of the $x^2-y^2$, $xz$, $yz$ and $xy$ orbits relative to the $z^2$ orbit under compression is shown in

Figure 6. Under compression, energies of the *yz* and the *xy* orbits at Fe1 (4*a*) and Fe2 (8*f*) sites moves upward while energies of the $z^2$, $x^2$-$y^2$, and *xz* orbits stay almost invariant. Then electrons on the *yz* and *xy* orbits transfer to the $z^2$, $x^2$-$y^2$, and *xz* orbits in order to minimize the total energy in $Fe_4O_5$ crystal. In contrast, the orbital-energy differences of the Fe3 (4*c*) site increase much slower under compression, which leads to the slow magnetic moment collapse at the Fe3 (4*c*) site in Fig. 3.

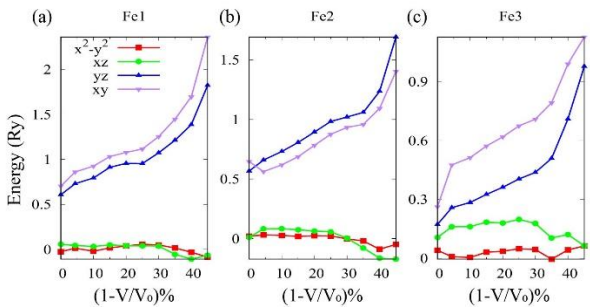

Fig. 6 (Color online.) Pressure dependence of energy difference among the $x^2$-$y^2$, *xz*, *yz* and *xy* orbits relative to the $z^2$ orbit at Fe1 (4*a*), Fe2 (8*f*) and Fe3 (4*c*) at 300 K in panels (a)-(c) respectively.

## CONCLUSION

Our study shows that $Fe_4O_5$ of $CaFe_3O_5$-type is always metallic from the ambient pressure to high pressures. Under compression the three non-equivalent iron atoms in $Fe_4O_5$ exhibit site-dependent collapse of magnetic moments. The magnetic moments of iron at Fe1 (4*a*-site) and Fe2 (8*f*-site) collapse from about 4.0 $\mu_B$/Fe at the ambient pressure to about 1.0 $\mu_B$/Fe at the 45% compression endpoint. But the magnetic moment of iron at Fe3 (4*c*-site) decreases much slower since it has the strongest electron correlations among the three non-equivalent iron sites in $Fe_4O_5$. By further calculation of the electron occupancies and the energy levels of the five Fe-3*d* orbits, we prove that the site-selective magnetic moment collapse in $Fe_4O_5$ results from the shift of orbital energies and the charge transfer among these orbits under compression. Our calculations indicate high conductivity and possible volume collapse in $Fe_4O_5$ at the extreme conditions, which will affect the dynamics in Earth's deep interior such as the energy exchange and the mass partition at the core-mantle boundary.

## ACKNOWLEDGEMENT

A.Q. Y., Q.Y. Q. and P. Z. acknowledge the support of National Natural Science Foundation of China No. 11604255 and the Natural Science Basic Research Program of Shaanxi No. 2021JM-001. Y.T. Z. acknowledges the support of National Key R & D Program of China under Grant No. 2019YFA0404900, the Science Challenge Project under Grant No. TZ2016005. This work is also supported by the HPC platform of Xi'an Jiaotong university.

# Supplementary Information

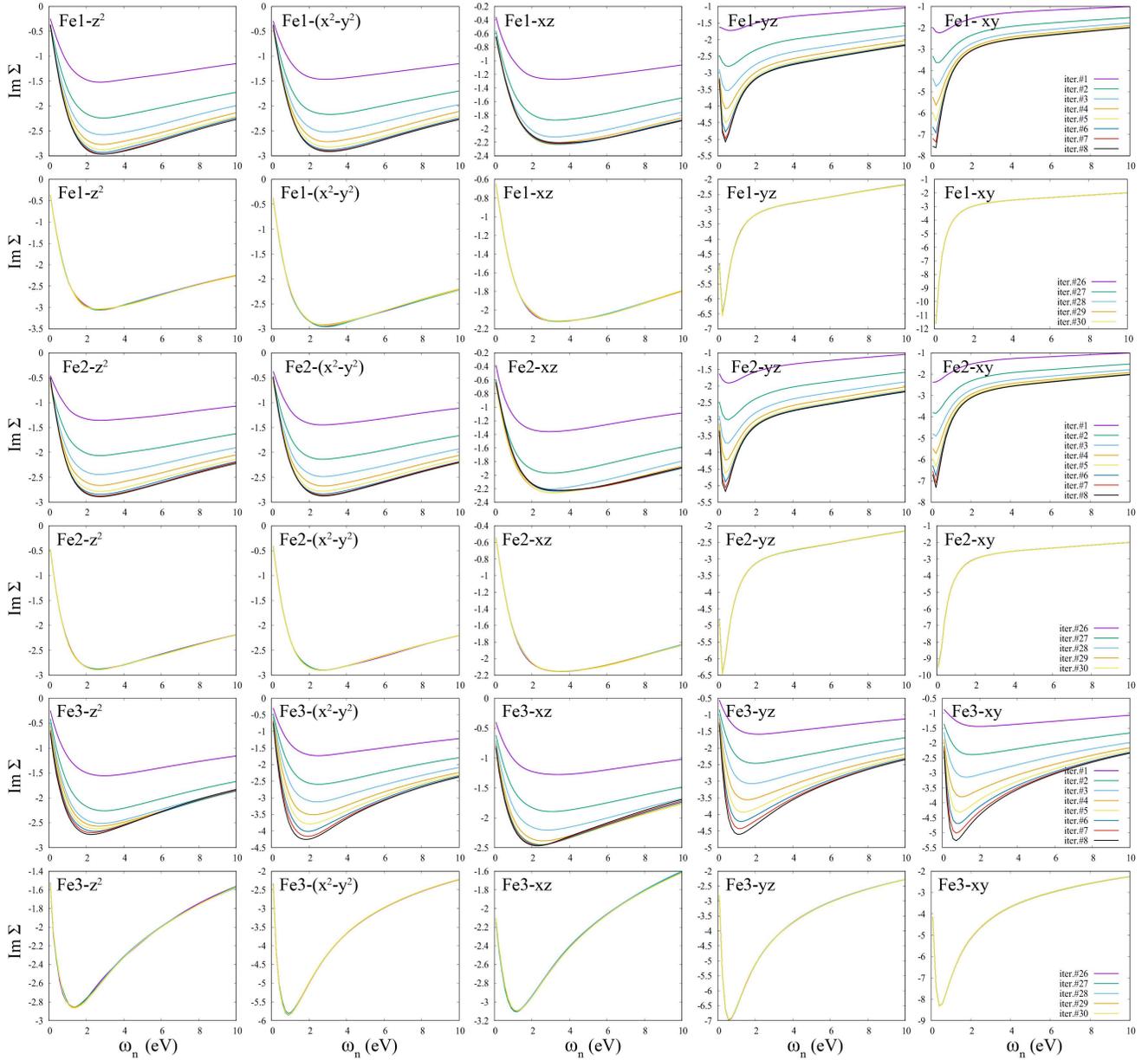

Fig. S1 (Color online.) Evolution of the self-energy with respect to the DMFT iterations for the five Fe-3$d$ orbits of the three iron atoms at 300 K and at the ambient pressure. The self-energy is almost converged after the 8$^{th}$ iteration. The curves of the last five iterations overlap, which indicates that the self-energy has converged. The $x$-axis is at the Matsubara frequency.

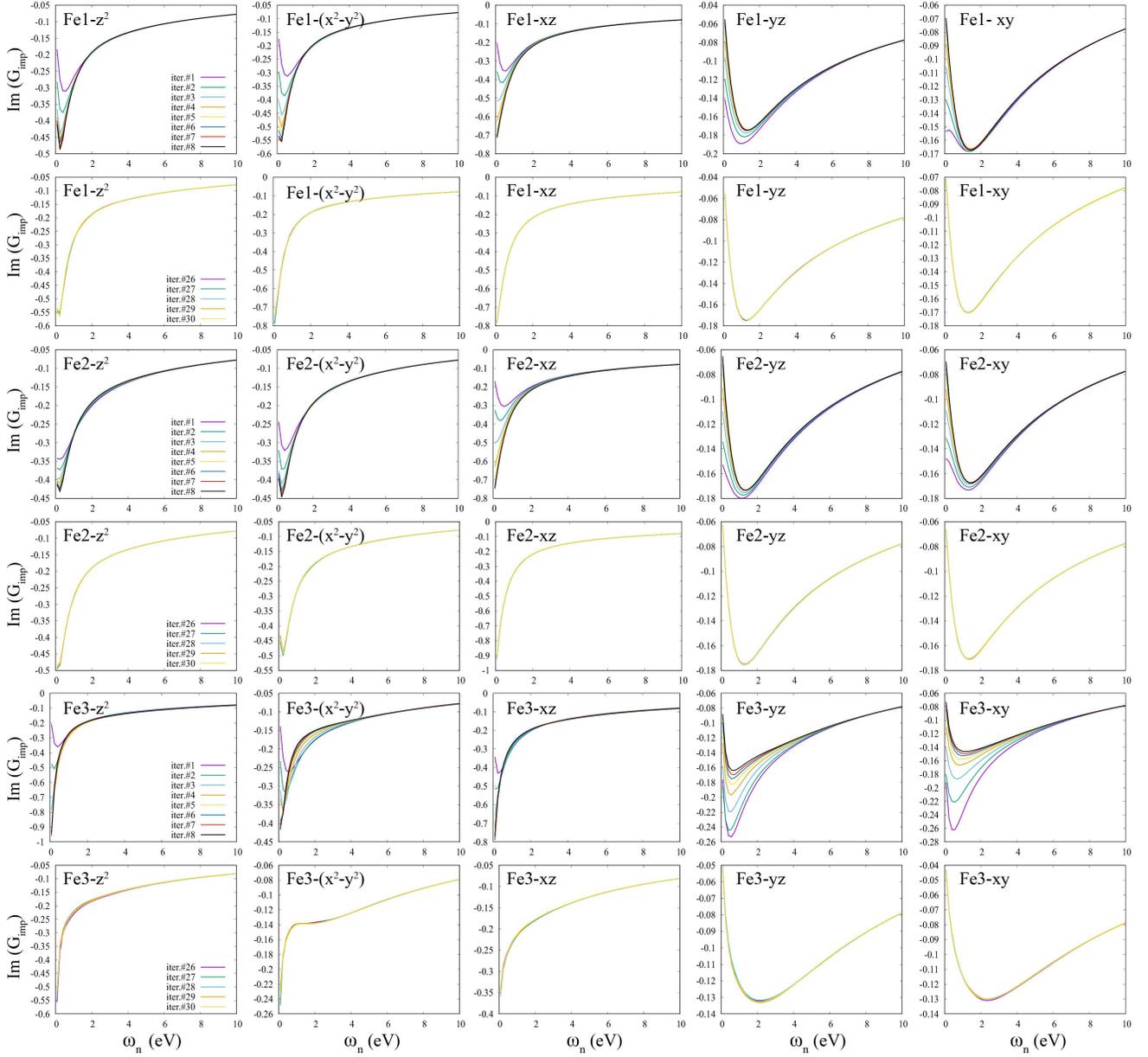

Fig. S2 (Color online.) Evolution of the impurity Green's function for the five Fe-3$d$ orbits of the three iron atoms with respect to the DMFT iterations at 300 K and at the ambient pressure. The Green's function is almost converged after the 8[th] iteration. The curves of the last 5 iteration overlap, which indicates that the local Green's function has converged. The $x$-axis is at the Matsubara frequency.

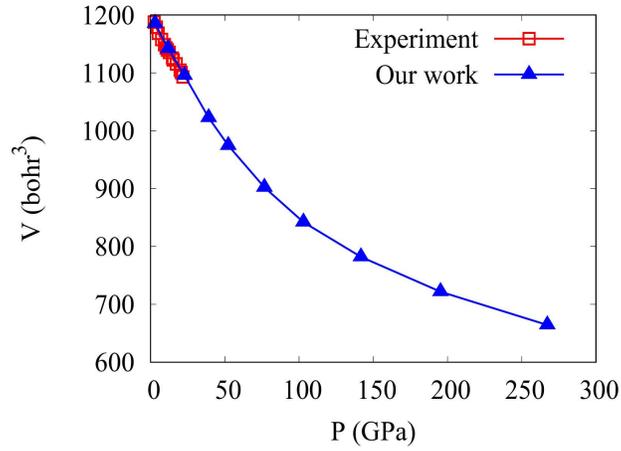

Fig. S3. The pressure-volume relationship of $Fe_4O_5$ at 300 K by fitting to the Birch-Murnaghan *equation of state*. The open squares below 30 GPa correspond to the powder x-ray diffraction data by Lavina *et al.* [15].

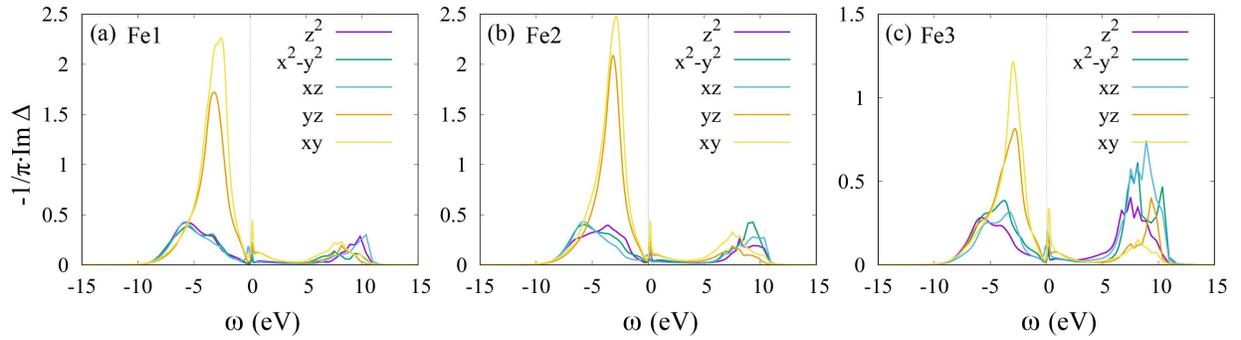

Fig. S4 (Color online.) The hybridization functions for the five Fe-*3d* orbits of the three iron atoms at 300 K and at the ambient pressure. The *x*-axis is at the real frequency.